\documentclass[a4paper,twoside]{jpconf}
\usepackage{graphicx}
\usepackage{fancyhdr}
\begin{document}
\pagestyle{fancy}
\rhead
[
\setlength{\fboxrule}{0.00pt}
\setlength{\fboxsep}{0.0mm}
\fbox{
\parbox{0.75\linewidth}{
{\footnotesize
K. P. N. Murthy, {\it Metropolis and Wang-Landau Algorithms}
, in BRNS School on
} 
}}
]{\thepage}

\lhead[\thepage]
{
\setlength{\fboxrule}{0.00pt}
\setlength{\fboxsep}{0.0mm}
\fbox{
\parbox{0.75\linewidth}{
{\footnotesize
{\it Computational Methodologies across Length Scales,} 
Aug. 28 - Sep. 9, 2017, BARC
}
}}
}  

\chead{}\lfoot{}\cfoot{}\rfoot{}

\hfill
\setlength{\fboxrule}{0.65pt}
\setlength{\fboxsep}{0.5mm}
\fbox{
\parbox{.70\linewidth}{
\begin{center}
{\footnotesize
BRNS School on 
{\it Computational Methodologies across Length Scales,}

August 28 - September 09, 2017, BARC, Mumbai
} 
\end{center}
  }}
\vglue -15mm
\title{Metropolis and Wang-Landau Algorithms}
\author{K. P. N. Murthy}
\address{Chennai Mathematical Institute (CMI),\\
H1, SIPCOT IT Park, Sirisur, Kelambakkam,\\
 Chennai 603 103 Tamilnadu\\[1mm] and\\[1mm]
 Manipal Center for Natural Sciences (MCNS),\\
Manipal University, Manipal 576 104 Karnataka}
\ead{kpnmurthy@cmi.ac.in
}
\begin{abstract}
Metropolis algorithm has been extensively employed for
simulating a canonical ensemble and  
estimating macroscopic properties of a closed system at 
any  desired temperature. A mechanical property, 
like energy can be calculated  by averaging 
over a large number of micro states of the stationary Markov chain
generated by 
the Metropolis algorithm.  However thermal 
properties like entropy, and free energies  are not easily accessible. 
A method called {\it umbrella sampling}  
was proposed some forty years ago for this purpose. 
Ever since, umbrella sampling has undergone several metamorphoses and we have now 
multi canonical Monte Carlo, entropic sampling, 
flat histogram methods, Wang-Landau algorithm {\it etc.} 
In these talks I shall tell you of  Metropolis algorithm   for estimating 
mechanical properties and of Wang-Landau algorithm for estimating 
both mechanical and thermal 
properties of an equilibrium  system.
I shall make these lectures as pedagogical and self-contained as possible.
\end{abstract}
\section*{Some Preliminaries}
Statistical mechanics helps us go from 
the micro world of atoms and molecules obeying
laws of classical and quantum mechanics to the 
macro world of thermodynamics describing matter 
in bulk. In a single stroke, Ludwig Eduard Boltzmann (1844 - 1906) 
connected physics at the length scales of atoms and molecules to
phenomena on length scales of solids, liquids,  gases, polymers,
magnets, {\it etc.} The micro-macro
synthesis proceeds, very generally, along the following lines.  

First, we  identify  a random variable 
that corresponds to a thermodynamic property. 
The average of the random variable 
over a suitable and well defined statistical ensemble\footnote{{\it e.g.} 
Gibbs' ensembles : micro canonical for an isolated system;
canonical  for a 
closed system; and grand canonical  for an open system.

The notion of an ensemble came from James Clerk Maxwell (1831-1879).
A {\bf Maxwell's ensemble}  is a set,  whose elements are drawn 
from the micro states of the system 
under considerations. A micro state occurs in the ensemble several times.
Number of times it occurs divided by the size of the ensemble equal its probability. 
Thus, an ensemble contains, not only information 
about the micro states of the system, but 
also about their probabilities. 

Imagine now,  a collection of a large number of 
identical mental copies of the macroscopic system under consideration. 
They constitute a 
{\bf Gibbs'  ensemble}. Each member of the Gibbs' ensemble 
shall be in some micro state of the other. 
Different members can be in different micro states;
But all of them have the
same macroscopic properties. This is what we mean 
when we say they are 
identical mental copies of the system. When the number of elements 
in the Gibbs' ensemble is large
then the number of systems in a given micro state divided by the size of the ensemble 
will give the probability of the micro state.

Given the micro states and their probabilities, we can construct an ensemble.
This is what we do in Monte Carlo. Given an ensemble, 
we can calculate the probabilities of the micro states of the macroscopic system to which the 
ensembles belongs. 
This is what we do when we derive Boltzmann weight for the micro states 
employing  {\it the method of most probable distribution}  where we  construct  
a Canonical ensemble. See any standard text book on statistical mechanics 
{\it e.g.} the book written by Pathria \cite{RKP}.   
}
gives the value of the 
thermodynamic property. As an example, consider 
internal energy\footnote{
{\bf Internal Energy and the First Law of Thermodynamics :} 
In thermodynamics, internal energy is defined 
completely in terms of work done in adiabatic processes :
Select a reference point $O$  in the thermodynamic phase plane. 
Define a function $U$ as follows. 
Assign an arbitrary
value to $U(O)$.  
Consider a point $A$.  
Measure or calculate work done in an adiabatic process that takes 
the system from $O$ to $A$.
Then define : $U(A)=U(O)+W_{O\to A}^A.$ 
The superscript $A$ tells that 
the process considered is adiabatic. Employ the convention~: 
{\it work done on the system is positive and work done by the system is
negative}. By considering adiabatic processes 
we can define $U$ at all points on the phase plane.
If there exists a point, say $B$,  which is not 
accessible adiabatically from $O$ then consider
an adiabatic process that takes the system from $B$ to $O$ for 
purpose of defining $U$ :
$U(B)=U(O)-W^A_{B\to O}.$
Then consider an arbitrary process from $C$ to $D$.
Let $W$ be the work done and  $\Delta U=U(D)-U(C).$ 
Then, $\Delta Q=\Delta U-W$ is called heat and this is a statement of 
the first law of thermodynamics. Heat is the difference 
between actual work and adiabatic work. 
Thus the first law of thermodynamics establishes the 
mechanical equivalence of heat. 
As an off-shoot of the first law of thermodynamics 
we get to  
define a thermodynamic property called the internal energy, denoted by the 
symbol $U$.}
of a thermodynamic system. This property is 
usually 
denoted by the symbol $U$. Corresponding to this property, we have, in statistical mechanics,  
energy $E$ - the kinetic energy and the interaction energy of the atoms 
and molecules of the macroscopic object. A numerical value for $E$ 
can be assigned to each micro state\footnote{For example three positions $(q_1,q_2,q_3)$ 
and three momenta $(p_1,p_2,p_3)$ 
are required to specify a single point particle. 
For $N$ particles, we need a string of 
$6N$ numbers and this string denotes a micro state 
of the macroscopic system of $N$ particles.
$$E=\frac{1}{2m}\sum_{i=1}^{3N}p_i^2+V(q_1,q_2,\cdots ,q_{3N}).$$ 
The energy consists of  the kinetic energy and potential energy. 
Note that energy is defined for each micro state.  
For a macroscopic system of say $N$ Ising spins, we have $2^N$ micro states 
since each Ising spin can be in 
either "up" ($S_i=+1$) state or "down" ($S_i=-1$) state. 
$$E=-J\sum_{\langle i,j\rangle} S_i\ S_j,$$
where $S_i$ is the spin  at lattice site $i$ and $J\ >\ 0$ measures the strength of 
spin-spin interaction. Spins on nearest neighbour lattice sites interact. The sum runs over all 
pairs of nearest neighbour spins}
 of the macroscopic system. 
The value of $E$ will fluctuate when the equilibrium system goes 
from one micro state to another. These fluctuations are an integral  
part of an equilibrium description. 
The average of $E$ 
gives  the  internal energy~: $\langle E\rangle=U$, 
and the fluctuations are proportional to the   heat capacity\footnote{called
fluctuation dissipation theorem relating equilibrium fluctuations to
response of the system to small perturbation.} : 
$\langle E^2\rangle-\langle E\rangle ^2\propto C_V$. 
The symbol $\langle (\bullet)\rangle$ denotes 
averaging  of the property $(\bullet)$  over the chosen 
ensemble. 
\section*{Energy}
The computation of average energy is now straight forward~:
Generate a canonical ensemble employing, for example, 
 Monte Carlo method based on Metropolis algorithm; a simple arithmetic
average of $E$ over a Monte Carlo sample of reasonably large size,  gives the required answer. 
The statistical error associated
with the finite-sample  average can also be calculated from the 
data obtained in the simulation. Such a neat 
computational scheme is possible because a numerical value 
for energy can be assigned to each 
micro state of the macroscopic system. 
\section*{Entropy}
How does one calculate entropy ? 

We can not assign a numerical value for entropy to any single micro state. 
Entropy is a property
that belongs collectively to all the micro states. While energy is a 
{\it private}  property (of each 
micro state) entropy is  a {\it social} or a {\it pubic}  property, see below.

Let 
$\Omega=\{ {\cal X}_\nu\ :\nu=1,\ 2,\ \cdots ,\widehat{\Omega}\}$ 
denote the set
 of  micro states of an equilibrium
 system; the micro states are discrete, distinct  and finite in number.
$\{ p({\cal X}_\nu)\ :\ \nu=1,\ 2,\ \cdots ,\widehat{\Omega}\}$
are their 
probabilities. We use 
\lq\  script ${\cal X}$\ \rq\  to denote micro states of the system and 
 \lq\ roman ${\rm X}$\ \rq\   to denote micro states of an ensemble or of a Monte Carlo sample 
or of a Markov chain. The Boltzmann-Gibbs-Shannon entropy is given by 
$$S=-k_B\sum_{\nu=1}^{\widehat{\Omega}} p({\cal X}_\nu)\ \ln p({\cal X}_\nu).$$
In the above, $k_B$ is the Boltzmann 
constant\footnote{$k_B=1.38064852 \times 10^{-23}$ joules/kelvin
is called 
the Boltzmann constant. It helps us convert energy 
measured in units of kelvin to energy in units of 
joule.}.
\section*{Entropy of an Isolated System}
For an isolated system  the micro states  are equally 
probable\footnote{we call it {\it ergodicity}; it is an hypothesis; the entire edifice of 
statistical  mechanics is built on this hypothesis.}. We have, 
$$
p({\cal X}_\nu)=\frac{1}{\widehat{\Omega}(E,V,N)}\ \forall\ \nu .
$$
$\widehat{\Omega}(E,V,N)$ is the number of micro states of the isolated system of  
of $N$ particles, confined to a volume $V$, and with a  fixed total energy of $E$.
For an isolated system the expression  for entropy simplifies to 
$S(E,V,N)=k_B\ln \widehat{\Omega}(E,V,N).$
\section*{Entropy of a Closed System}
For a closed system at temperature\footnote{In thermodynamics,
 temperature is defined as 
${\displaystyle T=(\partial U/\partial S)_{V,N}.}$}
  $T=1/[k_B\beta]$, we have
$$p({\cal X}_\nu)=\frac{1}{Q}\ \exp[-\beta E({\cal X}_\nu)],\ \  
{\rm where}\ \    
Q=\sum_{\nu=1}^{\widehat{\Omega}} \exp[-\beta E({\cal X}_\nu)],$$
is called the canonical partition function. 
\section*{Entropy of an Open System}
For an open system we have
$$p({\cal X}_\nu)=\frac{1}{{\cal Q}}\ \exp[-\beta\{ E({\cal X}_\nu)-\mu N({\cal X}_\nu)\}],\ 
{\rm where}\  
{\cal Q}(T,V,\mu)=\sum_{\nu=1}^{\widehat{\Omega}}\  \exp[-\beta\{E({\cal X}_\nu)-
\mu N({\cal X}_\nu)\}],$$
is the grand canonical partition function. In the above,
$\mu$ is the chemical potential\footnote{the chemical potential
gives the change in energy upon addition of a single 
particle keeping the entropy and volume of the 
system at a constant value.  words 
${\displaystyle \mu=(\partial U/\partial N)_{S,V}.}$} 
of the system, and $N({\cal X}_\nu)$ is the 
number of particles in  the system when it is in micro state ${\cal X}_\nu$.   

Our aim is to simulate physical processes occurring in an equilibrium system
and assemble a large number of micro states 
consistent with the 
given probabilities. To this end, we start with an arbitrary initial micro 
state ${\rm X}_0(\in \Omega)$; then, employing Metropolis 
rejection algorithm \cite{MRRTT}
we
generate a Markov chain\footnote{\label{FN-Markov-Chain} {\bf Markov Chain} : Consider a sequence
of micro states visited by the 
system at discrete times starting from $X_0$ at time $0$. Let us denote the 
sequence by 
${\rm X}_0\to {\rm X}_1\to\cdots\to {\rm X}_{n-1}\to {\rm X}_n,$
where the subscript denote the discrete time index. Our interest is to calculate the 
joint probability  of the sequence.   
From Bayes' theorem we have,
$$P(\ {\rm X}_n,{\rm X}_{n-1},\cdots {\rm X}_1,{\rm X}_0)=
P(\ {\rm X}_n\ \vert\  {\rm X}_{n-1},{\rm X}_{n-2}\cdots {\rm X}_1,{\rm X}_0\ )\times  
P(\ {\rm X}_{n-1},{\rm X}_{n-2}\cdots {\rm X}_1,{\rm X}_0\ ).$$
If 
$P\left(\  {\rm X}_n\ \vert\  {\rm X}_{n-1},{\rm X}_{n-2},
\cdots {\rm X}_1,{\rm X}_0\ \right)=
P\left(\  {\rm X}_n\ \vert\  {\rm X}_{n-1}\ \right),$ 
then  ${\rm X}_0\to {\rm X}_1\to\cdots {\rm X}_{n-1}\to
{\rm X}_n$ is a Markov chain : The future depends only on the present
and not on the past. Thus, once the present is specified, the future is independent of the past. 

Under Markovian condition, the expression for the joint probability 
of the chain of micro states, simplifies to
\begin{eqnarray*}
P(\ {\rm X}_n,{\rm X}_{n-1},\cdots {\rm X}_1,{\rm X}_0\ ) & =  &
P(\ {\rm X}_n\ \vert\  {\rm X}_{n-1}\ )\times P(\ {\rm X}_{n-1},{\rm X}_{n-2}\cdots {\rm X}_1,{\rm X}_0\ ),
\end{eqnarray*}
\begin{eqnarray*}
    &=& P(\ {\rm X}_n\ \vert\  {\rm X}_{n-1}\ )\times P(\ {\rm X}_{n-1}\ \vert\  {\rm X}_{n-2}\ )\times  P(\ {\rm X}_{n-2},{\rm X}_{n-3}\cdots {\rm X}_1,{\rm X}_0\ ),\\
    &=& \cdots \cdots\cdots ,\\
    &=&P(\ {\rm X}_n\ \vert\  {\rm X}_{n-1}\ )\times P(\ {\rm X}_{n-1}\ \vert\  {\rm X}_{n-2}\ )\times 
                          \cdots \times P(\ {\rm X}_1\ \vert\  {\rm X}_0\ )\times P(\ {\rm X}_0\ ).
    \end{eqnarray*}
    Since we are interested in equilibrium properties
    we consider
    a sequence of states visited by an equilibrium system :  The conditional probability, 
    $P(\ {\rm X}_n\ \vert\  {\rm X}_{n-1}\ )$ is independent of the time 
    index. In other words 
    $$P(\ {\rm X}_n={\cal X}_\mu\ \vert\  {\rm X}_{n-1}={\cal X}_\nu\ ) = W_{\mu,\nu},$$ 
    and this quantity is 
    independent of time. We call it time homogeneous Markov chain. Once we know the 
    transition probability matrix $W$  
     and initial probabilities of all the micro states,  
     we can calculate the probability  of any given Markov Chain.
     The transition probability matrix $W$ is a square matrix of size 
     $\widehat{\Omega}$. We have $$0\le W_{\mu,\nu}\le 1\ \ \forall\ \ \mu,\nu \ \ \ \ {\rm and}\  
     \ \ \ \sum_{\mu=1}^{\widehat{\Omega}} \ W_{\mu,\nu}=1\ \forall\ \nu.$$
     $W$ is called Markov matrix or stochastic matrix. Its elements are all between 
     zero and unity; The elements of each column add to unity. Besides, 
     if the elements of each row also add to unity, then we have a doubly stochastic matrix.  
},
$${\rm X}_0(\in \Omega)\to {\rm X}_1(\in \Omega)\to {\rm X}_2(\in \Omega)\to \cdots\to 
{\rm X}_i(\in \Omega)\to {\rm X}_{i+1}(\in \Omega)\to \cdots $$
\section*{Metropolis Rejection Algorithm}
Let us say we have simulated the Markov chain upto ${\rm X}_i\in\Omega$ 
starting from ${\rm X}_0\in\Omega$. 
Thus the current micro state is ${\rm X}_i$. Let $p_i=p({\rm X}_i)$ denote its probability.
We make a
small random change in the current micro state and construct a 
trial 
micro state\footnote{For example if we are simulating an Ising spin system, select randomly an 
Ising spin from the current spin configuration (micro state) and flip it to construct 
a trial spin configuration. If we are simulating a collection of particles, then select a particle 
randomly and change its there position coordinates and three momentum  coordinates by
small random amounts to construct a trial micro state.}
${\rm X}_t\in\Omega$. Let
$p_t=p({\rm X}_t)$ denote its probability. Calculate
$p={\rm minimum} \left( 1,\ p_t/p_i\right).$
Then, the next micro state in the Markov chain is given by,
\begin{eqnarray*}
{\rm X}_{i+1}=\left\{\begin{array}{llc}
 {\rm X}_t & {\rm with\  probability}\ & p\\[2mm]
                 {\rm X}_i & {\rm with\ probability}\  & 1-p
                 \end{array}\right.\end{eqnarray*}
The implementation of the above goes as follows :
\begin{enumerate}
\item[$\bullet$]
Generate a random number\footnote{employ the random number generator available
in your computer.
The (pseudo) random  numbers are real numbers independently and 
uniformly distributed between zero and one. Random number generation and testing are 
non-trivial tasks and they constitute highly  specialized areas of research. 
Most Monte Carlo practitioners
are not aware of the subtleties and difficulties associated with random number generation
employing deterministic algorithms and testing of the generated random numbers for 
randomness. We take the available  random generator and use it as a 
black box.} 
uniformly distributed between zero and unity. Denote it by the symbol $\xi$.
\item[$\bullet$]
If $\xi\ \le\ p$, then accept the trial state and advance the Markov chain to ${\rm X}_{i+1}={\rm X}_t$.
\item[$\bullet$]
If not, reject the trial state and advance the Markov chain to 
 ${\rm X}_{i+1}={\rm X}_i$.
 \item[$\bullet$]
Repeat the process on  ${\rm X}_{i+1}$ to get ${\rm X}_{i+2}$; and so on.
\end{enumerate} 
 Generate a long Markov chain.
The asymptotic part of the chain shall contain micro states belonging to the ensemble 
characterized by the 
probabilities $\{ p({\rm X}_\nu)\ :\ \nu=1,\ 2,\ \cdots\}$. 
\section*{Important Properties of Metropolis Algorithm}
\begin{enumerate}
\item[$\bullet$]
Metropolis algorithm demands only a 
 knowledge of the ratio of probabilities
of  two micro states belonging to $\Omega$.
We should know this ratio for all possible pairs of micro states
of $\Omega$.
This implies that  we need to know $\{ p({\cal X_\nu}) : \nu=1,2,\cdots\widehat{\Omega}\}$ 
only up to a normalization 
constant.
It is precisely because of this reason we are able to simulate a closed system, 
since we need to know only the Boltzmann weight $\exp[-\beta E({\cal X})]$ of 
each micro state; we need not have any knowledge what so ever of the 
canonical partition function.
\end{enumerate}
\begin{enumerate}
\item[$\bullet$] 
Metropolis algorithm  obeys 
balance condition\footnote{\label{FN-ME-Balance}
We consider time homogeneous Markov chain, see footnote (\ref{FN-Markov-Chain}). 
 Let $P({\cal X}_j,n)$ be the probability
for the system to be in micro state ${\cal X}_j$at discrete time $n$. 
Let $W_{i,j}$ denote the probability for transition from micro state 
${\cal X}_j$ to micro state ${\cal X}_i$ in one time step. We have 
$W_{i,j}=P\ (\ {\cal X}_i\ \vert\  {\cal X}_j\ ),$ 
the conditional probability that the system is in micro state 
${\cal X}_i$ at any instant of time  given  it was in 
micro state ${\cal X}_j$ at the previous instant of time.
The probabilities obey the equation given below.
\begin{eqnarray*}P({\cal X}_i;n+1) &=& \sum_{j\  :\  j\ne i}\ P({\cal X}_j,n)\ W_{i,j}+P({\cal X}_i,n)\ W_{i,i}\end{eqnarray*}
We have $\sum_i W_{i,j}=1\ \forall \ j.$  Therefore, $W_{i,i}=1-\sum_{j\ :\ j\ne i} W_{j.i}.$ 
We can write the above equation as
\begin{eqnarray*}
P({\cal X}_i;n+1) &=& \sum_{j\ne i}\ P({\cal X}_j,n)\ W_{i,j}+\ P({\cal X}_i,n) 
\left( 1-\sum_{j\ :\ j\ne i} W_{j.i}\right)\\[1mm]
            & = & P({\cal X}_i,n) +\sum_{j\ne i}\ \left[ P({\cal X}_j,n)\ W_{i,j}- P({\cal X}_i,n)W_{j,i}\right]\end{eqnarray*}
            {\bf Balance Condition} : 
When the system equilibrates we have 
$P({\cal X}_i,n+1)=P({\cal X}_i,n)=p({\cal X}_i)\ \forall\ i.$
Therefore we have 
$${\sum_{j}\ \big[ \ p({\cal X}_j)\ \times\  W_{i,j}\ -
\  p({\cal X}_i)\ \times\  W_{j,i}\ \big] = 0
   }.$$ This  is  called the balance condition which ensures that the Markov chain eventually 
   equilibrates.}.
The balance condition tells  that the Markov chain shall 
converge, definitely,  to an   invariant probability distribution. 
\item[$\bullet$]
Metropolis algorithm obeys a stricter condition called detailed 
balance\footnote{\label{FN-ME-Detailed-Balance}
{\bf Detailed Balance :} Look at the balance condition given 
toward the end of footnote \ref{FN-ME-Balance}
as  a sum over $j$ for each $i$. 
We can make a stricter demand that each term in the sum be zero. Then we 
 get the detailed balance condition : 
$$p({\cal X}_j)\ \times\ W_{i,j} \ =\   p({\cal X}_i)\ \times\ W_{j,i}\ \ \ \ \forall\ i,j=1,2,\cdots ,\widehat{\Omega}.$$
It is quite easy to show that the Metropolis rejection 
algorithm  obeys detailed balance condition. I leave this as an exercise for you.        
 }.
The consequences of this are two fold.
\begin{enumerate}
\item[(i)] Detailed balance ensures the Markov chain  
converges  to  
an equilibrium ensemble consistent with the given  probability weights of the micro states :
Boltzmann weights for canonical ensemble; and Gibbs weights for grand canonical
ensemble; {\it etc.}
We can choose the nature of the equilibrium state.
\item[(ii)] Detailed  balance ensures that the  Markov chain    
is reversible;
hence it is most suited
for describing an equilibrium system\footnote{By observing an equilibrium 
system we can not tell which direction time flows.
Both directions are  equally probable 
and equally unverifiable. 
Consider a Markov chain of micro states visited by an equilibrium system : 
${\rm X}_0\to {\rm X}_1\to\cdots {\rm X}_n\to {\rm X}_{n+1}\to\cdots {\rm X}_M. $
The transition probabilities are given by 
$W_{i,j}=P({\rm X}_n={\cal X}_i\vert {\rm X}_{n-1}={\cal X}_j)$
 
At discrete time $M$ let us reverse the Markov chain and get 
${\rm X}_M\to {\rm X}_{M-1}\to \cdots {\rm X}_{n+1}\to {\rm X}_n\to\cdots {\rm X}_1\to {\rm X}_0.$
A little thought will tell you the above is also a Markov chain : 
for, the future depends only on the present and not on the past for a Markov chain,  
Hence once the present is specified 
the future is independent of the past. Past is independent of the future which renders 
the time reversed  chain,  Markovian. 
Let us denote the transition probability matrix of the time reversed 
chain by the symbol $W^R$. We have
\begin{eqnarray*} 
W^R_{i,j} = P({\rm X}_{n}={\cal X}_i\vert {\rm X}_{n+1}={\cal X}_j)
           =  \frac{P({\rm X}_{n}={\cal X}_i,\   {\rm X}_{n+1}={\cal X}_j)}{p({\cal X}_j)}
    &=&  \frac{P({\rm X}_{n+1}={\cal X}_j\vert {\rm X}_n={\cal X}_i)\ p({\cal X}_i)}{p({\cal X}_j)}\\
          & = & \frac{W_{j,i} \ p({\cal X}_i)}{p({\cal X}_j)}
          \end{eqnarray*}
The condition for reversibility is $W^R_{i,j} = W_{i,j}$ : The transition probability matrix
should be the sane for both  Markov chains - the time forward and the time reversed.
Hence on the left hand side of the above equation replace $W^R_{i,j}$ by $W_{i,j}$ and 
reorganize. Then  the condition for reversibility reads as,
$$W_{i,j}\ p({\cal X}_j)\ =\ W_{j.i}\ p({\cal X}_i). $$ 
We recognize this as   detailed balance, see footnote (\ref{FN-ME-Detailed-Balance}). 
Thus a Markov chain of micro states of an equilibrium system 
obeys detailed balance condition and hence is reversible;
}; 
for, no matter what kind of observations you make on an equilibrium system, 
you can not 
tell which way  time moves.
Equilibrium  is a time-reversal invariant state. Detailed 
balance captures
this subtle property. 
\end{enumerate}
\end{enumerate}
\section*{Estimation of Averages and Statistical Errors}
Generate a Markov chain until it equilibrates\footnote{calculate the moving average of energy. 
When it stabilizes to  a constant value but for some small statistical fluctuations, we can say the 
system has equilibrated.}. Continue the Markov chain and collect a reasonably large number of micro
states $\{ {\rm X}_i \ :\ i=1,\ 2,\ \cdots M\}$ from the equilibrated Markov chain. 
Let $O$ be a property of interest and $O({\rm X})$ its value 
when the system is in micro state ${\rm X}$. Then the  Monte Carlo estimate of the 
property $O$ is given by\footnote{We reserve the symbol
 $\langle O\rangle$ to denote the 
unknown exact value of the canonical ensemble average of the property $O$ formally given by 
$$\langle O\rangle =
  \frac{1}{Q}
\sum_{\nu=1}^{\widehat{\Omega}} O({\cal X}_\nu)\exp[-\beta E({\cal X}_\nu))];\ \ \ \ \ 
Q = {\sum_{\nu=1}^{\widehat{\Omega}}\exp[-\beta E({\cal X}_\nu))]}.
$$
},
\begin{eqnarray*}
\overline{O}_M = \frac{1}{M}\sum_{i=1}^M\ O({\rm X}_i);\ \ \ \ \ 
{}^{\ {\rm Limit}}_{M\to\infty}\ \  \overline{O}_M  = \langle O\rangle .
\end{eqnarray*} 
A little thought will tell you that the quantity $\overline{O}_M$
is a random variable. It will fluctuate from one realization of a Monte Carlo 
sample  to another. 

What is the nature of 
these fluctuations ? 

The Central limit theorem\footnote{{\bf Central Limit Theorem}: 
Let $X_1,X_2,\cdots ,X_M$ be identically distributed independent random variables
with finite mean,  $\mu$  and finite variance, $\sigma^2$. Let $Y=(X_1+X_2+\cdots +X_M)/M$. 
The central limit theorem CLT)
says that $Y$ is a Gaussian with mean $\mu$ and variance $\sigma^2/M$ 
when $M\to\infty$. CLT is a glorious
culmination of a series of studies starting with the Chebyshev 
inequality, see {\it e.g.} \cite{WF,AP} :  A single number 
randomly sampled from 
a distribution, with finite mean $\mu$,  and finite variance, $\sigma^2$ can 
fall out side the interval $\mu\pm k\sigma$ with a probability 
not more  than $1/k^2$.
Then came several laws of large numbers and these led eventually to 
the Central Limit Theorem (CLT),  see any standard text book,  
{\it e.g.} \cite{WF,AP}
 on probability theory and stochastic processes to know more on these issues.}
(CLT)
  tells that the quantity $\overline{O}_M$
is a Gaussian random variable when $M$ is  large. 
The average of 
the Gaussian  is $\langle O\rangle$ and its variance is 
$\sigma^2/M$, where $\sigma^2=\langle O^2\rangle-\langle O\rangle ^2$.
A possible statement  of the statistical error associated 
with the Monte Carlo estimate $\overline{O}_M$ is obtained 
from the following considerations. 

Take  a Gaussian random variable with mean $\zeta$ 
and standard deviation $\Sigma$. The area under 
the Gaussian\footnote{$$\frac{1}{\Sigma\sqrt{2\pi}}\ \int_{\zeta-\Sigma}^{\zeta+\Sigma}
 dx\ \exp\left[ - \frac{(x-\zeta)^2}{2\Sigma^2}\right]=0.682695$$}
between $\zeta-\Sigma$ and $\zeta+\Sigma$ is $0.682695$. This means  that 
with $68.27\%$ confidence, you can say that a randomly sampled number from 
the Gaussian shall lie between $\zeta-\Sigma$ and $\zeta+\Sigma$. 
The one-sigma confidence interval provides a neat quantification of the 
statistical error associated with Monte Carlo estimates, see below.

We calculate the second moment,  
\begin{eqnarray*}
\overline{O}^2_M = \frac{1}{M}\sum_{i=1}^M O^2({\rm X}_i);\ \ \ \ \  
  {}^{\ {\rm Limit}}_{M\to\infty}\ \  \overline{O}^2_M  =   \langle O^2\rangle. 
\end{eqnarray*}
From the calculated values of the first and second moments we estimate the variance as,
\begin{eqnarray*}
\sigma^2_M = \overline{O}^2_M-(\overline{O}_M)^2.\ \ 
\left( \sigma^2 = \langle O^2\rangle -\langle O\rangle ^2 =
{}^{\ {\rm Limit}}_{M\to\infty}\ \sigma^2_M  \right).
\end{eqnarray*}
We can now calculate the one-sigma confidence interval; we 
quote the Monte Carlo result as
$\overline{O}_M\pm \sigma_M/\sqrt{M}.$ 
The above means that with $0.6827$ probability we can expect the Monte Carlo estimate
$\overline{O}_M$  to lie in the one sigma interval around $\langle O\rangle$; {\it i.e.} to
lie between $\langle O\rangle -\sigma_M/\sqrt{M}$ and $\langle O\rangle+\sigma_M/\sqrt{M}$.

The statistical error decreases with increase of $M$.
This is indeed a desirable property. This tells us,
 atleast in principle, we will get things right  
if $M$ is sufficiently large.  
Usually we would be interested in comparing  our Monte Carlo 
predictions with 
experiments. Hence we can take the Monte Carlo sample size to be 
  large enough  to  ensure that the statistical error is less that the 
experimental error bar. 

However, notice the statistical error decreases 
painfully  slowly with  the  sample size. 
The decrease is logarithmically slow : 
to better the results  by one extra decimal 
accuracy  we need to increase the  sample size
a hundred fold. Often this  would prove to be an exercise
in futility; for, the computing time is
linear in $M$. 

We need  variance reduction devices 
that significantly reduce the fluctuations without 
affecting the averages.
Importance sampling is a variance reduction device. 
It helps us sample micro states from
important regions of the sample space {\it e.g.} micro states with high Boltzmann 
weights. Notice a randomly selected micro state 
would be, most likely, of high energy\footnote{entropy increases  with energy.}, 
hence of low Boltzmann weight; its contribution 
to the partition sum would be negligible. In fact the Metropolis algorithm  
is an importance sampling device. 
I am not going to talk of importance sampling or of other variance reduction 
techniques;  those interested can consult for example 
\cite{KPN-1,KPN-2,MI}.

Instead, in what follows, I am going to investigate the nature of the 
invariant distribution  of the Markov chain of micro states whose probabilities are 
inversely proportional to the density of states : 
micro states of high entropy region 
have low probabilities; and those of low entropy region
have high probabilities. This kind of prescription 
does not describe any physical system.
Nevertheless constructing a Markov chain with 
these probabilities for the micro states, 
has certain advantages and  this will become clear in the sequel.  
\section*{Markov Chain with Flat Energy Histogram}
Consider  a system with micro states 
$\Omega = \{ {\cal X}_\nu\ :\ \nu=1,2,\cdots ,M\}$. 
Let $\widehat{\Omega}(E)$ denote its density of states. 
For purpose of illustration we assume that the density 
of states is known.
Let ${\cal X}_\mu \in \Omega$ and $ E_\mu = E ({\cal X}_\mu )$. 
We prescribe  
$P({\cal X}_\mu)\ \propto\ 1/\widehat{\Omega}(E\mu).$
Let me emphasize two points, at the risk of being repetitive,  before we proceed further : 
\begin{enumerate}\item[$\bullet$] We do not know the density of states
before hand\footnote{ 
After all, if we know the density 
of states then we can make an estimate of all the properties of the system employing the 
well developed machinery of thermodynamics and statistical mechanics. There would arise
no  compulsive need for a Monte Carlo simulation. We may  still decide to carry out 
 Monte Carlo simulation, assemble an 'entropic' ensemble,  
and  extract physical quantities employing 
un-weighting and re-weighting techniques. I shall tell you of this later in my talk}. 
\item[$\bullet$] There is no physical system for which 
the probability of a micro state is inversely proportional to the density of states
\footnote{The set $\Omega$ shall contain all the micro states of the "un-physical" 
system. These micro states 
can be of different energy. Let us group them in terms of their energies.
Then we can say all the micro states of a group are equally probable and this probability 
is given by the inverse of the density of states at that group energy. 
Each group would then constitute a micro canonical  ensemble.}.
\end{enumerate} 
Nevertheless we shall consider  Monte Carlo simulation of such an un-physical 
system employing 
Metropolis algorithm and investigate the invariant probability density 
of the Markov chain it generates. 

Let ${\rm X}_i$ be the current micro state in the Markov chain 
and $E_i=E({\rm X}_i)$ its energy.
We have $p_i=p({\rm X}_i)\propto 1/\widehat{\Omega}(E_i)$. 
Let ${\rm X}_t$ be the  trial state and 
$E_t=E({\rm X}_t)$ its energy. 
We have $p_t=p({\rm X}_t)\propto 1/\widehat{\Omega}(E_t).$  
The probability of acceptance of the 
trial micro state is then given by
\begin{eqnarray*}
p &=& {\rm minimum}\left( 1, \frac{p_t}{p_i}\right)=
{\rm minimum}\left( 1,\frac{\widehat{\Omega}(E_i)}{\widehat{\Omega}(E_t)}\right)
\end{eqnarray*}
Note that if the trial micro state belongs to a lower entropy region it gets accepted
with unit probability; however  if it belongs to higher entropy region 
its acceptance probability is less than unity.
Thus the algorithm pushes the Markov chain preferentially toward low entropy region.
This preference cancels statistically exactly the natural tendency of randomly sampling 
of trial micro states from high entropy region. As a result  the Markov chain shall have equal 
number of micro states in equal regions of energy. In other words the energy histogram 
of the visited micro states shall be flat.

 Thus the Markov chain visits all regions of energy 
 with equal ease. It does not see any 
 energy barriers, insurmountable or otherwise, that 
 might be  present in the system under 
 investigation. This is a huge advantage because 
 there are indeed  energy barriers 
 that emerge at temperatures close to the first 
 order phase transition 
 and which are responsible for  
 super critical slowing of the dynamics. Also glassy systems 
 have free energy profile with  
 numerous ups and downs. Though we get an un-physical ensemble as a 
 result employing inverse of the density of states in 
 a Markov chain Monte Carlo method based on Metropolis rejection,
 there seem to be certain  desirable properties for the ensemble. 
 Of course we do not know yet the density of states. 
 Perhaps it is a good idea to investigate 
 further  and invent methods that  
 that help obtain the density of states. May be if we 
 embark on such an enterprise we may  have to 
   to forgo  the comforts of Markov Chain methodology and of
   the detailed balance present in  the Metropolis rule. 
   But then, we shall get easy access to entropy
    and other thermal properties,
   which eluded the Markov chain Monte Carlo practitioners. 

 What is it that renders calculation of entropy a difficult task ?  
  To answer this question we have to realize 
  that the usefulness of the Monte Carlo methods considered upto now, 
is tied crucially to our ability 
to assign a numerical value of the property $O$ 
to every micro state of the system.
Consider estimating a property like entropy. We can not assign
a numerical value for entropy to any single micro state of the system. 
All the micro states collectively own entropy.
Hence thermal properties in general and entropy in particular are not 
 easily accessible.
  
For computing thermal properties we need to go beyond Boltzmann Monte Carlo methods.    
That  non-Boltzmann 
sampling can provide a legitimate and perhaps superior alternative to Boltzmann methods has
been recognized even during the very early days of Monte Carlo practice, see 
{\it e.g.} \cite{Fosdick} and to these issues we turn our attention, below.   

Torrie and Valleau \cite{TV} were, perhaps, the first to propose 
a non-Boltzmann algorithm to 
calculate the thermal properties. 
Their method called umbrella sampling has since undergone 
a series of metamorphoses. We have the multi-canonical Monte Carlo of 
Berg and Neuhaus \cite{BN}, entropic sampling of Lee \cite{L} and 
the algorithm of Wang and Landau \cite{WL}.
We describe below the Wang-Landau algorithm.
\section*{Wang-Landau Algorithm}
Wang and Landau \cite{WL} proposed an algorithm to estimate  iteratively 
the density of states of the system. The algorithm is described below. 

At the beginning of the simulation, define a function $g(E)$ 
and set it to unity for all $E$. 
Define also an histogram $H(E)$ and set 
it to zero for all $E$. Start with an arbitrary initial micro state ${\rm X}_0$. 
Let $E_0=E({\rm X}_0)$ be its energy.
Update $g(E)$ and $H(E)$ as follows :
$$
g(E_0) = g(E_0)\times \alpha;\ \ 
H(E_0) = H(E_0)+1.
$$
Here $\alpha$ is the Wang-Landau factor and we take $\alpha=e^1=2.7183$ 
in the first  iteration. Generate a chain of micro states 
$${\rm X}_0\to {\rm X}_1\to \cdots \to {\rm X}_i\to{\rm X}_{i+1}\to\cdots$$ 
as per the algorithm described below.  

Let ${\rm X}_i$ be the current micro state. Construct a trial
micro state ${\rm X}_t$. We need to decide whether to accept the 
trial state for advancing the chain. We take a decision on the basis 
of the $g(E)$ updated at the end of the previous step in which we selected
the micro state ${\rm X}_i$. Let $E_i=E({\rm X}_i)$ and $E_t=E({\rm X}_t)$.
We have $$p_i\propto \frac{1}{g(E_i)}\  
{\rm and}\  p_t\propto \frac{1}{g(E_t)}.$$ Define
$$p={\rm minimum}\left(1,\frac{p_t}{p_i}\right)=
{\rm minimum}\left(\frac{g(E_i)}{g(E_t)}\right).$$ 
The next micro state in the chain
is
\begin{eqnarray*}{\rm X}_{i+1} =\left\{\begin{array}{lcl}
                            {\rm X}_t  & {\rm with\  probability} & \ p\\[3mm]
                            {\rm X}_i  & {\rm with\ probability} & 1-p
                            \end{array}\right.\end{eqnarray*} 
                            Once ${\rm X}_{i+1}$ is selected, the function $g(E)$ and 
                            the histogram $H(E)$ are updated.   
Carry out the 
simulation of the  chain of micro states until the energy histogram  becomes 
flat over, at least,  a small range of energy. This constitutes one Wang-Landau iteration.
 
Note that the density-of-state-function $g(E)$ is updated at every step and   
the updated function is employed for decision making, from  the very next step. 
As a result 
the chain of micro states generated, is not Markovian. 
The probability of transition 
between two micro states at any time step  in the chain 
depends on how many times the chain  has visited these two micro states
in its past. The transition from present to future depends on the entire past. 
Hence we shall refer to the sequence of micro states  
as simply a chain and not prefix it with the adjective "Markov". 

At the end of the first Wang-Landau iteration,
change $\alpha$ to $\sqrt{\alpha}$. Reset $H(E)$ to zero for all $E$; 
but continue with  
$g(E)$. Carry out the second Wang-Landau iteration. 
The histogram would spread out and 
would at the same time  become flatter over a wider range of energy. 

Upon further iterations the value of alpha will move closer and 
closer to unity. For example, after some twenty five iterations we shall have 
$\alpha=1+3\times 10^{-7}$. The histogram of energy would become flat 
at least over the range of energy of interest 
after a few Wang-Landau iteration runs.
 
The flatter the histogram,  closer would be $g(E)$ to
the true but unknown density of states $\widehat{\Omega}(E)$. 
We take $g(E)$ obtained at the end of the 
last iteration - the one which generates a  reasonably flat energy histogram, 
 as an estimate of $\widehat{\Omega}(E)$, the true density of states. 
 
 We can define a suitable criteria
 for measuring the flatness of the histogram. 
 For example we can consider  the histogram to be  flat if 
 the smallest and largest entries  do not differ from each other by say more 
 than say ten percent. Depending upon the requirement of accuracy 
 and the availability  of 
 computing resources, we can relax or tighten the flatness criterion.
 
 There is no hard and fast rule about either the choice of 
 the initial value of the Wang-Landau factor 
 or about how it decreases to unity  from one iteration to the next. 
 The choice of  $\alpha=\alpha_0=e^1$ at the beginning of the first iteration 
 and the square-root rule of decrease, were recommended by 
 Wang and Landau\cite{WL}. In principle,  
 $\alpha_0$ can be any real number greater than unity and it should 
 decrease,  preferably monotonically,  to  
 unity. Some authors,  see {\it e.g.} \cite{JSM,PCABD},  
 have  found it advantageous 
vary $\alpha$ non-monotonically at least initially.
The important point is any choice of variation of $\alpha$ that 
flattens the  histogram would serve the purpose. 
In a sense the histogram provides a diagnostic tool
with which you can monitor whether you are doing things right or wrong. 
The flatness of the histogram tells 
you how close has  the density of states converged to its true value. 

 The Wang-Landau algorithm estimates  the density of states only 
 upto a normalization constant. In other words the micro canonical entropy is estimated 
 only upto an additive constant. This is quite adequate 
  since we need to calculate only change in entropy rather than absolute 
 entropy in almost all applications. 
 
In principle we can stop here. Once we know the density of 
states then we can employ the machinery of thermodynamics 
and know everything else about the system. 
\section*{Entropic Ensemble} 
Alternately, we can employ the converged density of states in a production 
run and generate a large ensemble of micro states. The sequence of micro states generated
 in the production run  constitute  a legitimate 
Markov chain, obeying detailed balance. However the invariant 
probabilities  are un-physical : the probability of a micro state ${\cal X}$ is inversely proportional  
the density of states at $E=E({\cal X})$.  
The Markov chain obeys detailed balance  and hence 
convergence to the desired ensemble, though unphysical, 
is guaranteed.  

Let us call the set of micro states generated in the production run as  an 
an entropic ensemble or Wang-Landau ensemble. By employing 
 un-weighting and 
re-weighting techniques\footnote{\label{FN-importance-sampling}
Let me explain un-weighting and re-weighting
 in a simple manner\cite{KPN-1,KPN-2}. Let
$x$ be a random variable and $f(x)$ its probability density. Let $h(x)$ be some function 
of $x$. The $f$-ensemble average of $h$ is formally expressed as,
$$\langle h\rangle_f = \int_{-\infty}^{+\infty}\ dx\ h(x)\ f(x),$$
Let $g(x)$ be a density function. Let us  generate an ensemble
$\Omega_g=\{ x_i\ :\ i=1,2,\ \cdots M\}$ by random sampling from $g(x)$. 
Our aim is to make an estimate of $\langle h\rangle _f$ employing the set $\Omega_g$.
Consider the following.
\begin{eqnarray*}
\langle h\rangle_f &=& \int_{-\infty}^{+\infty}\ dx\ h(x)f(x)\ 
                 = \int_{-\infty}^{+\infty}\ dx\ h(x)\frac{f(x)}{g(x)}g(x)\ 
                 =\left\langle\ \  h\ (1/g)\ f\ \ \right\rangle_g
                 \end{eqnarray*}
                 The above is an exact result. The left hand side is an $f$ ensemble average of
                 $h$. The right hand side is a $g$ ensemble average of $h$ un-weighted by $1/g$ and re-weighted by $f$. The implementation goes as follows.
 $$\langle h\rangle _f = {}^{\ {\rm Limit}}_{M\to\infty}\   \frac{1}{M}\ \sum_{i=1}^M\  h(x_i)\ \times 
 \frac{1}{g(x_i)}\ \times\  f(x_i);\ \ \ \ \ x_i\ \in\ \Omega_g.$$
 }
we can  make from the entropic ensemble, statically reliable 
estimates of physical quantities. 

In what follows I  shall show how to convert the entropic ensemble to
a micro canonical ensemble 
and to a canonical ensemble.
\section*{Entropic Ensemble $\to$ Micro Canonical Ensemble}
Let $\{ {\rm X}_i\ :\ i=1,2,\cdots ,M\}$ denote a set of $M$ micro states belonging to the entropic ensemble. These micro states have been sampled from a probability 
distribution $$p({\rm X}_i)\propto 1/g(E({\rm X}_i)).$$ Hence we first carry out 
un-weighting, see footnote (\ref{FN-importance-sampling}) :
$$W({\rm X}_i)=\frac{1}{1/g(E({\rm X}_i))}= g(E({\rm X}_i).$$
Note that the micro states of the entropic ensemble are not necessarily of the same energy. 
In fact the ensemble contains equal number of micro states in equal regions of energy - in other words
the energy-histogram is flat. For a micro canonical ensemble all micro states are of the same energy and 
are equally probable. Hence  the re-weighting 
factor is $1\times \delta (E-E({\rm X}_i))$; the 
delta function  ensures 
that we assemble only those micro states with the desired energy.  Thus we have
$$W({\rm X}_i)=g(E({\rm X}_i))\ \delta(E-E({\rm X}_i)).$$  

Let $O({\rm X}_i)$ be the value of a property when the system is in micro state ${\rm X}_i$. The micro canonical ensemble average of $O$ is given by,
\begin{eqnarray*}
\langle O\rangle _{\mu C}(E) &=& {}^{\ {\rm Limit}}_{M\to\infty}\frac{
             \sum_{i=1}^M\ O({\rm X}_i)g(E({\rm X}_i))\delta (E({\rm X}_i)-E)}
                         {\sum_{i=1}^M g(E({\rm X}_i))\delta (E({\rm X}_i)-E)}
                         \end{eqnarray*}
                         In the above we have taken $E$ as the energy of the isolated system 
                         described by the  micro canonical ensemble.

Thus weighted averaging over micro states of given energy belonging to the 
                         {\it un-physical} entropic ensemble equals averaging over a {\it physical} 
                         micro canonical ensemble modeling an isolated system.

  \section*{Entropic Ensemble $\to$ Canonical Ensemble}                       

The un-weighting factor remains the same as the one derived 
 for converting entropic ensemble to micro canonical ensemble. 
         The re-weighting factor however is the Boltzmann weight. Thus
                $$W(C_i)=g(E({\rm X}_i))\times\exp[-\beta E({\rm X}_i)].$$
                       All the micro states of the entropic ensemble contribute
                       to the canonical ensemble average.
                         
                         The canonical ensemble average of $O$ is  given  by
                         \begin{eqnarray*}
\langle O\rangle _{C} &=& {}^{\ {\rm Limit}}_{M\to\infty}\frac{
                         \sum_{i=1}^M\ O({\rm X}_i)g(E({\rm X}_i))\exp[-\beta E({\rm X}_i)]}
                         {\sum_{i=1}^M g(E({\rm X}_i))\exp[-\beta E({\rm X}_i)]}
                         \end{eqnarray*}
                Thus the weighted 
                         average over the {\it unphysical} entropic ensemble is 
                         equivalent to average over 
                         a {\it physical} canonical ensemble modeling a closed system.
                        
      From one single ensemble of micro states we can calculate averages 
 over a large number of distinct canonical ensembles at different 
  temperatures. This is a huge advantage especially for problems in which we need 
          the properties on a fine grid of temperatures in the neighbourhood of a
    phase transition.   
\section*{End Note} 

I have talked about Metropolis algorithm to sample 
micro states from a given ensemble, 
physical or otherwise. If sampling is done from a 
physical ensemble we call it Boltzmann 
Monte Carlo. Boltzmann sampling has been 
eminently successful 
for estimating mechanical properties
like energy. The reason is simple. 
A value for a mechanical property 
can be assigned to each micro state. 
 
 However Boltzmann sampling is 
 quite clumsy when it comes to estimating 
 thermal properties 
 like entropy and free energies. 
 The clumsiness owes its origin to the fact that a numerical value for  
 entropy can not be assigned to any single  micro state. 
 All the micro states, collectively, own entropy. 
 Entropy is a property of an ensemble and 
 not of any single  micro state. 
 This problem about estimating entropy 
 was recognized  even in the 
 early days of Monte Carlo practice by   
 Torrie and Valleau\cite{TV}; they  invented umbrella sampling which  
  addresses these issues. 
 Umbrella sampling has since inspired and 
 given rise to a whole lot  of non-Boltzmann methods; 
 the  latest  to arrive  
 is the method of  Wang and Landau\cite{WL}. 
 I have told you of the basic idea behind  
 Wang-Landau algorithm and described 
  how to implement it
 on a practical problem. 
   
The take-home-message is that 
non-Boltzmann Monte Carlo methods 
are as good as Boltzmann methods, if not more,  
for calculating mechanical properties. 
Besides, they  provide reliable estimates of thermal properties,  
not easily accessible to Boltzmann Monte Carlo methods. 

I must quickly add that all is not cozy about 
Wang-Landau algorithm.   
There are  issues and there are difficulties. 
A typical Monte Carlo aficionado, 
see {\it e.g.} \cite{WJ}, 
does not feel comfortable since 
the algorithm  does not obey detailed balance; 
in fact, the chain generated is 
not  Markovian.
What guarantees convergence 
of $g(E)$ to $\widehat{\Omega}(E)$ ? 

Also the  algorithm  
performs  poorly  
on systems with continuous degrees of freedom.  
There is a slowing down of dynamics but now  
due to entropy barriers.
These and related  problems 
have  attracted the attention of 
several authors, see {\it e.g.} \cite{JSM, PCABD,SBM,ZB,TD}
and remedies have been suggested.
But in my opinion no satisfactory solution has yet emerged. 
All the remedies suggested seem ad-hoc.
 
There are also issues  about  
error - both systematic and statistical - associated 
with the computed density of states. 
How does one translate the non-flatness of the 
energy histogram to error bars in the estimated density of states ?
After all, the pride of a Monte Carlo practitioner lies often in 
his ability to compute averages but also   
 associated statistical errors.  But then we do not know
 how to calculate Monte Carlo error bars 
in Wang-Landau simulation. 

I hope  these and other issues would get
resolved satisfactorily soon and let me end the talk with this 
optimistic note. In case you want to discuss 
further  on issues raised  in this talk,
do not hesitate to get in touch with me at k.p.n.murthy@gmail.com  (.)  
\section*{Acknoledgement}
I thank Swpan K Ghosh, D K Maity and Ashok Arya,   
for the invitation.   
My special  thanks to {\bf Ashok Arya},  
for the hospitality  and for simply being what he is - a wonderful human being
and a great organizer.  
I owe a special word of  thanks to {\bf Manoj Warrier}, 
for coding the Wang - Landau algorithm,   on a  toy problem  involving 
coin tossing.  I must also thank him  for conducting the hands-on
session on Monte Carlo methods,  with  gusto. 
I must thank all the participants 
of the School for the questions they raised, 
for the comments they made, and for the
several lively discussions on and off the business hours. 
  

\section*{References}
\begin{thebibliography}{100}

\bibitem{RKP}
 R K Pathria, {\it Statistical Mechanics}, 
Second Edition, Butterworth - Heinemann (1996).

\bibitem{MRRTT}
N.Metropolis, A.W.Rosenbluth, M.N.Rosenbluth, A.H.Teller, and E.Teller,
{\it Equation of State Calculations by Fast Computing Machines},
 Journal of  Chemical  Physics  {\bf 21}
1087 (1953); see also G.Bhanot, {\it The Metropolis Algorithm}, Reports of Progress in Physics {\bf 51}, 429 (1988).



\bibitem{WF} W.Feller, {\it An Introduction to Probability Theory and Applications I and II},
John Wiley (1968)

\bibitem{AP} A.Papoulis, {\it Probability Theory, 
Random Variables, and Stochastic Processes} 
McGraw Hill (1965)

\bibitem{KPN-1}
K.P.N.Murthy, {\it Monte Carlo Methods in Statistical Physics}, Universities Press (2004)

\bibitem{KPN-2} 
K.P.N.Murthy, {\it Monte Carlo : Basics}, Report ISRP - TD-3, Indian Society for Radiation Physics, Kalpakkam Chapter (2000); see arXiv:cond-mat/014215v1,  12 Apr. 2001.


\bibitem{MI}
E.J.McGrath, and D.C.Irving, 
{\it Techniques for Efficient 
Monte Carlo Simulation} Volume III : Variance
Reduction, Report SAI - 72 - 590 - LJ , 
Office of the Naval Research, 
Department of the Navy, Arlington,
Virgenia 22217 USA (March 1973)

\bibitem{Fosdick}
L.D.Fosdick, {\it Monte Carlo Computation on the Ising Lattice} in
 Methods of Computational Physics, Vol. 1, Editor B Adler, p.245 (1963).

\bibitem{TV}
G.M.Torrie, and J.P.Valleau,
{\it Non-physical sampling distributions 
in Monte Carlo free-energy estimation: 
Umbrella sampling},
 Journal of  Computational  
 Physics  {\bf 23} 187 (1977).

\bibitem{BN}
B.A.Berg, and T.Neuhaus,
{\it Multi canonical ensemble: 
A new approach to simulate 
first-order phase transitions},
Physical  Review  Letters {\bf 68}, 9 (1992).

\bibitem{L} J.Lee,  
{\it New Monte Carlo algorithm: Entropic sampling}, 
Physical  Review  Letters  {\bf 71}, 211 (1993);
{\it Erratum},  {\bf 71}, 2353 (1993)

\bibitem{WL} F.Wang, and D.P.Landau,
{\it Efficient, Multiple-Range Random Walk Algorithm to 
Calculate the Density of States,} 
Physical  Review  Letters {\bf 86}, 2050 (2001).

\bibitem{JSM} D.Jayasri, V.S.S.Sastry, 
and K.P.N. Murthy, {\it Wang-Landau Monte Carlo
simulation of isotropic-nematic transition in 
liquid crystals}, 
Physical Review E {\bf 72}, 36702 (2005).

\bibitem{PCABD} P.Poulin, F.Calvo, R.Antoine, 
M.Broyer, and, P.Dugord, 
{\it Performances of Wang-Landau algorithms 
for continuous systems},
Physical Review E {\bf 73}, 56704 (2006).

\bibitem{WJ} 
W.Janke, {\it Monte Carlo Simulations in 
Statistical Physics – From Basic
Principles to Advanced Applications}, in:
Order, Disorder and Criticality:  
Advanced Problems of Phase Transition Theory, 
Vol. 3  (edited by Y. Holovatch), 
World Scientific  (2012)pp. 93-166.

\bibitem{SBM}
B.J.Schulz, K.Binder, M.M\"uller, and D.P.Landau, 
{\it Avoiding Boundary Effects in Wang-Landau Sampling}
Physical Review E {\bf 67}, 67102 (2003).

\bibitem{ZB} C.Zou, and R.N.Bhatt, 
{\it Understanding and Improving the Wang-Landau
Algorithm}, Physical Review E {\bf 72},
25701(R) (2005). 

\bibitem{TD} A. Tr\"ster, and C. Dellago, 
{\it Wang-Landau Sampling with Self Adaptive
Range}, Physical Review E {\bf 71},
66705 (2005),

\end{thebibliography}
\end{document}